\documentclass[amsmath,prb,twocolumn,superscriptaddress]{revtex4}

\usepackage{color}
\usepackage{amsmath}
\usepackage{amsfonts}
\usepackage{graphicx}

\begin{document}


\author{Massimiliano Esposito}
\affiliation{Center for Nonlinear Phenomena and Complex Systems,
Universit\'e Libre de Bruxelles, CP 231, Campus Plaine, 
B-1050 Brussels, Belgium}
\author{Michael Galperin}
\email{migalperin@ucsd.edu}
\affiliation{Department of Chemistry \& Biochemistry, 
University of California at San Diego, La Jolla CA 92093, U.S.A.}

\title{A self-consistent quantum master equation approach to molecular transport}
\date{\today}


\begin{abstract}
We propose a self-consistent generalized quantum master equation (GQME) to describe electron transport through molecular junctions. In a previous study [M.Esposito and M.Galperin. Phys. Rev. B \textbf{79}, 205303 (2009)], we derived a time-nonlocal GQME to cure the lack of broadening effects in Redfield theory. To do so, the free evolution used in the Born-Markov approximation to close the Redfield equation was replaced by a standard Redfield evolution. In the present paper, we propose a backward Redfield evolution leading to a time-local GQME which allows for a self-consistent procedure of the GQME generator. This approach is approximate but properly reproduces the nonequilibrium steady state density matrix and the currents of an exactly solvable model. The approach is less accurate for higher moments such as the noise.
\end{abstract}

\maketitle

\section{\label{intro}Introduction}
Recent advances in the experimental capabilities for constructing molecular junctions and measuring their response to external perturbations create 
new challenges for an adequate theoretical description of open quantum systems far from equilibrium.\cite{NitzanRatner, LindsayRatner, our_jpcm, our_review} In molecular junctions, most of the interesting applications are concerned with either strongly correlated or resonant tunneling regimes (or both). In these situations, conventional perturbation theory and/or effective mean-field theories (e.g. Hartree-Fock, GW, DFT) are inapplicable and may lead to unphysical predictions.\cite{Millis}

In molecular junctions, molecules are sensible to processes such as oxidation, reduction, and excitations. This makes the description of transport in the language of the many-body states of the isolated molecules a convenient tool for molecules weakly coupled to the contacts. Furthermore, sophisticated quantum chemistry methods use many-body states to describe molecular electronic structures.\cite{Jensen} The dressed state representation,\cite{Nitzanbook} often employed in quantum chemistry, is another example of many-body states formulation. The standard nonequilibrium Green function (NEGF) approach,\cite{Danielewicz, RammerSmith, HaugJauhobook} is formulated in the language of elementary excitations and is therefore not well suited for a many-body states description. A suitable alternative, the Hubbard NEGF approach, \cite{Sandalov, Fransson, Hubbard} still displays inconsistencies for low level of approximation that need to be resolved (see Ref.~\onlinecite{Hubbard} for a discussion). 

The simplest approach to transport at the molecular state level is the Redfield quantum master equation (QME) originally developed in the context of NMR\cite{Redfield57} and later in many other fields\cite{Nitzanbook, Breuer02,GaspNaga99} including transport through quantum junctions.\cite{Yan05,Esposito06,vonOppen09} It is derived using the Born-Markov approximation in order to get a closed evolution equation for the reduced density matrix. The rotating wave (or secular) approximation (RWA)\cite{Breuer02,Esposito06} is often invoked when the molecular levels are well separated so that the molecule Bohr frequencies evolve fast compared to the relaxation time scale induced by the coupling to the contact. This leads to a QME which has a simple physical interpretation in the molecular eigenbasis. Indeed, populations satisfy a rate equation with rates given by Fermi golden rule (Pauli equation) and the coherences are each independently exponentially damped. As a result, in nonequilibrium steady-states, all coherence effects (in the molecular eigenbasis) are lost. In its general form, the Markovian Redfield equation predicts steady-state coherences that can be inaccurate and even sometimes lead to unphysical results.\cite{Tannor97a,Tannor97b,Esposito06,Fleming09} Furthermore, broadening effects are totally absent from the Redfield description.\cite{Wacker05,Neuhauser04,Wegewijs08,EspoGalpPRB}. In order to overcome some of these difficulties, we proposed in our previous work\cite{EspoGalpPRB} a generalized QME capable of predicting the broadening of the molecular levels in an approximate way. The key idea was to replace the assumption of free molecular evolution inside the kernel (which is second order in the molecule-contact interaction) used in the Born-Markov approximation by the Redfield evolution. The resulting equation was however non-local in time. Here we extend our consideration by proposing a kind of time-reversed Redfield evolution that leads to a time local generalized QME. This procedure is rewarding because it allows to formulate a practical self-consistent scheme to calculate the generator of the QME. 

The structure of the paper is the following. Section \ref{model} introduces a model of molecular junction. In section \ref{qme}, we briefly review our generalized QME derived earlier and introduce the self-consistent scheme. Section \ref{num} presents numerical examples and compares the results obtained within the scheme to a numerically exact approach. Conclusions are drawn in section \ref{conclude}. 


\section{\label{model}Model}
We consider a molecule ($M$) coupled to two metal contacts ($L$ and $R$) 
each at its own equilibrium. The Hamiltonian of the system is
\begin{equation}
 \label{H}
 \hat H=\hat H_M + \sum_{K=L,R}\left(\hat H_K + \hat V_{KM}\right)
\end{equation}
The contacts are assumed to be reservoirs of free charge carriers
\begin{equation}
 \label{HK}
 \hat H_K = \sum_{k\in K}\varepsilon_k\hat c_k^\dagger\hat c_k
\end{equation}
where $K=L,R$ and $\hat c_k^\dagger$ ($\hat c_k$) are creation 
(annihilation) operators for an electron in state $k$. 

The molecular Hamiltonian is represented in terms of the many-body 
states $\{|M\rangle\}$ of the isolated molecule
\begin{equation}
 \label{HM}
 \hat H_M = \sum_{M_1,M_2} |M_1\rangle H^{(M)}_{M_1M_2} \langle M_2|
 \equiv \sum_{M_1,M_2} H^{(M)}_{M_1M_2} \hat X_{M_1,M_2}
\end{equation} 
where $\hat X_{M_1,M_2}$ are projection (or Hubbard) operators.
In particular, in the eigenbasis representation, the molecular Hamiltonian reads
\begin{equation}
 \label{HMeig}
 \hat H_M = \sum_M E_M\hat X_{M,M}
\end{equation}
where $E_M$ are the eigenenergies.
Note that the molecular many-body states $|M\rangle$ are characterized by
all the relevant quantum numbers describing the state of an isolated molecule
or a dressed state.

The coupling between the molecule and the contacts is introduced in the 
usual way with hopping terms for the electrons moving from the contact to 
the molecule and vice versa
\begin{equation}
 \hat V_{KM} = \sum_{k\in K}\sum_\mathcal{M}\left(
 V_{\mathcal{M}k}\hat X_\mathcal{M}^\dagger\hat c_k + 
 V_{k\mathcal{M}}\hat c_k^\dagger\hat X_\mathcal{M} \right)
\end{equation}
Here
\begin{equation}
 \label{M}
 \mathcal M \equiv(N_f\, s_f,N_i\, s_i)
\end{equation}
is a transition from the molecular 
state $M_i\equiv N_i\, s_i$ to the molecular state $M_f\equiv N_f\, s_f$. 
The number of electrons on the molecule is $N_f=N_i-1$ and $s_i$ ($s_f$) is the set 
of all the quantum numbers characterizing the molecular state in the charging block 
$N_i$ ($N_f$), so that $\hat X_\mathcal{M}\equiv\hat X_{M_f,M_i}$.


\section{\label{qme}Generalized quantum master equation}
In this section we start by reviewing the derivation of the generalized QME
introduced in our previous work. We then introduce our new way of closing the 
master equation by using a time-reversed effective evolution, and
demonstrate how the resulting equation can lead to a self-consistent scheme to 
calculate the propagator of the molecular density matrix.


\subsection*{\label{derivation}Exact equation of motion}
We start by deriving exact equation of motion (EOM) for the quantity  
\begin{equation}
 \label{X21}
 \langle\hat X_{21}(t)\rangle \equiv 
 \mbox{Tr}\left[e^{+i\hat H(t-t_0)}\hat X_{21}e^{-i\hat H(t-t_0)}
                \hat\rho_0\right] \equiv
 \sigma_{12}(t)
\end{equation}
where $|1\rangle$ and $|2\rangle$ are many-body molecular states,
$t_0$ is a starting point of the evolution usually taken at the
infinite past $t_0\to-\infty$, and $\hat\rho_0$ is the initial 
density operator for the whole system (molecule and contacts).
We note that 
\begin{equation}
 \label{DM}
 \sigma_{12}(t) \equiv \langle 1|\,\mbox{Tr}_K\left[\hat\rho(t)\right]\,
 |2\rangle
\end{equation}
is a matrix element of the reduced density operator of the molecule
$\hat\sigma$ at time $t$ which is obtained by tracing out the contact
degrees of freedom from the full density matrix.

The Heisenberg EOM for the Hubbard operator in Eq.(\ref{X21}) yields
an {\em exact}\/ EOM for the reduced density matrix 
(see Ref.~\onlinecite{EspoGalpPRB} for details of derivation) 
which in molecular eigenbasis reads
\begin{align}
 \label{EOM}
 &\frac{d}{dt}\sigma_{12}(t) = -i\left(E_1-E_2\right)\sigma_{12}(t)
 -\sum_\mathcal{M}\sum_s\int_{-\infty}^t dt'
 \nonumber \\
 &i\left\{(-1)^{N_1-N_2}\left[\Sigma^{<}_{(N_1-1\, s,1)\mathcal{M}}(t-t')
  \langle\hat X_\mathcal{M}(t')\,\hat X^\dagger_{N_1-1\, s,2}(t)\rangle
 \nonumber \right.\right. \\ & \qquad\qquad \left.
 +\langle\hat X_{N_2-1\, s,1}(t)\,\hat X^\dagger_\mathcal{M}(t')\rangle 
 \Sigma^{<}_{\mathcal{M}(N_2-1\, s,2)}(t'-t)\right]
 \nonumber \\
 &\quad -\Sigma^{<}_{(2,N_2+1\, s)\mathcal{M}}(t-t')
  \langle\hat X_\mathcal{M}(t')\,\hat X^\dagger_{1,N_2+1\, s}(t)\rangle
 \nonumber \\ &\quad
  +\langle\hat X^\dagger_\mathcal{M}(t')\,\hat X_{N_2-1\, s,1}(t)\rangle
    \Sigma^{>}_{\mathcal{M}(N_2-1\, s,2)}(t'-t)
 \nonumber \\
 &\quad +\Sigma^{>}_{(N_2-1\, s,1)\mathcal{M}}(t-t')
   \langle\hat X^\dagger_{N_1-1\, s,2}(t)\,\hat X_\mathcal{M}(t')\rangle
 \nonumber \\ &\quad
  -\langle\hat X_{2,N_1+1\, s}(t)\,\hat X^\dagger_\mathcal{M}(t')\rangle
    \Sigma^{<}_{\mathcal{M}(1,N_1+1\, s)}(t'-t)
 \nonumber \\ & 
  -(-1)^{N_1-N_2}\left[\Sigma^{>}_{(2,N_2+1\, s)\mathcal{M}}(t-t')
   \langle\hat X^\dagger_{1,N_2+1\, s}(t)\,\hat X_\mathcal{M}(t')\rangle
  \right. \nonumber \\ &\left.\left. \qquad\qquad
  +\langle\hat X^\dagger_\mathcal{M}(t')\,\hat X_{2,N_1+1\, s}(t)\rangle 
   \Sigma^{>}_{\mathcal{M}(1,N_1+1\, s)}(t'-t) \right]\right\}
\end{align}
Here, $\mathcal{M}$ is defined in Eq.(\ref{M}),
$\Sigma^{>\,(<)}_{\mathcal{M}_1\mathcal{M}_2}$ are the greater 
(lesser) projections of a self-energy due to the coupling to the contacts
\begin{equation}
 \label{SE}
 \Sigma_{\mathcal{M}_1\mathcal{M}_2}(\tau_1,\tau_2) \equiv 
 \sum_{K=L,R} \sum_{k\in K} V_{\mathcal{M}_1k}\, g_k(\tau_1,\tau_2)\,
 V_{k\mathcal{M}_2}
\end{equation}
where $g_k(\tau_1,\tau_2)\equiv-i\langle T_c\hat c_k(\tau_1)\,
\hat c_k^\dagger(\tau_2)\rangle$ is the free electron Green function
($T_c$ is the contour ordering operator).
Dissipation matrix $\Gamma$ is introduced in a usual way
\begin{equation}
 \Gamma_{\mathcal{M}_1\mathcal{M}_2}(E) =
 i\left[\Sigma^r_{\mathcal{M}_1\mathcal{M}_2}(E)
       -\Sigma^a_{\mathcal{M}_1\mathcal{M}_2}(E)\right]
\end{equation}
and within wide band approximation assumed below is energy independent.

Eq.(\ref{EOM}) is {\em exact} but does not have a closed form in 
terms of the reduced density matrix $\sigma_{12}$. Indeed, its right 
hand side contains two-time correlation functions of Hubbard operators.
Attempt to write EOM for the latter would lead to expressions having 
in their right hand side correlation functions of higher order involving
both, molecular Hubbard operators and creation (annihilation) 
operators for the electrons in the contacts, at different times. 
This would generate an infinite hierarchy of EOMs. 


\subsection*{\label{GQME}Generalized time-nonlocal QME}
The usual strategy involves closing the system of equations by {\em approximately} 
representing high order correlation function in terms of correlation function(s) of 
lower order. In particular, closing Eq.(\ref{EOM}) for the case when $M$ are effective
single particle orbitals at the first step (i.e. reducing the two-time correlation
function in the right hand side to a single time average) is known as 
the generalized Kadanoff-Baym ansatz (GKBA).\cite{HaugJauhobook} 
In our previous work \cite{EspoGalpPRB}, 
we introduced the analogue of the GKBA in Liouville space 
where $M$ are molecular many-body states: 
\begin{align}
 \label{GKBA}
 &\langle\hat X_{12}(t)\hat X^\dagger_{34}(t')\rangle
 \approx 
 \nonumber \\ &
 i\sum_{M_1,M_2}\left[
 \mathcal{G}^r_{12,M_1M_2}(t-t')\langle\hat X_{M_1M_2}(t')\,
 \hat X^\dagger_{34}(t')\rangle 
 \right. \\ & \left. \qquad\quad 
 -\langle \hat X_{12}(t)\,
 \hat X^\dagger_{M_1M_2}(t)\rangle\mathcal{G}^a_{M_1M_2,34}(t-t')
 \right]
 \nonumber
\end{align} 
A similar expression holds for 
$\langle\hat X^\dagger_{34}(t')\, \hat X_{12}(t)\rangle$. 
Here $\mathcal{G}^{r\, (a)}$ are retarded (advanced) Green functions in 
Liouville space
\begin{align}
 \mathcal{G}^r_{12,34}(t) &\equiv -i\theta(t) \ll \hat X_{21}\hat I_B \rvert
 e^{-i\mathcal{L}t}\rvert\hat X_{43}\hat\rho_B^{eq}\gg 
 \nonumber \\
 &\equiv -i\theta(t)\ll\hat X_{21}\rvert e^{-i\mathcal{L}_{eff}t}
  \rvert\hat X_{43}\gg
 \\
 \mathcal{G}^a_{12,34}(t) &\equiv -i\theta(-t) \ll \hat X_{34}\hat I_B \rvert
 e^{i\mathcal{L}t}\rvert\hat X_{12}\hat\rho_B^{eq}\gg 
 \nonumber \\
 &\equiv i\theta(-t) \ll\hat X_{34}\rvert e^{i\mathcal{L}_{eff}t}
  \lvert \hat X_{12}\gg
\end{align}
where $\hat\rho_B^{eq}\equiv\hat\rho_L^{eq}\hat\rho_R^{eq}$
is the equilibrium density operator for the contacts
and $\hat I_B$ is unity operator in the contacts subspace.
Using (\ref{GKBA}) in (\ref{EOM}), leads to a time-nonlocal
generalized QME given by Eq.(35) in Ref.~\onlinecite{EspoGalpPRB}. 


\subsection*{\label{antiGQME}Generalized time-local QME}
We now propose an alternative way to close Eq.(\ref{EOM}) by reducing the two-time 
correlation function in the right hand side of (\ref{EOM}) to the later rather 
than earlier time. Indeed, since the whole system evolution is time-reversible 
(one has to be careful with spin and magnetic field, though), and since the 
reduction of a two-time correlation function to a single-time average is an 
approximation, the reduction of the correlation function to a later 
time does not a priori seem to be worse than to the earlier one.

The two-time correlation function can be {\em exactly}\/ expressed in 
Liouville space as
\begin{align}
 \label{corr}
 &\langle\hat X_{12}(t)\,\hat X^\dagger_{34}(t')\rangle =
 \nonumber \\ & \qquad
 \theta(t-t')\ll\hat X_{34}\hat I_B\rvert e^{-i\mathcal{L}(t'-t)}
    \lvert \hat\rho(t)\hat X_{12}\gg
 \\ & \quad
 + \theta(t'-t)\ll \hat X_{12}^\dagger\hat I_B\rvert e^{-i\mathcal{L}(t-t')}
    \lvert \hat X_{34}^\dagger\hat\rho(t')\gg
 \nonumber
\end{align}
Then using the projection superoperator introduced in Ref.~\onlinecite{EspoGalpPRB}
\begin{equation}
 \mathcal{P} = \sum_{M_1,M_2} \lvert\hat X_{M_1,M_2}\hat\rho_B^{eq}\gg\,
   \ll\hat X_{M_1,M_2}\hat I_B\rvert
\end{equation}
one gets the alternative ansatz
\begin{align}
 \label{anti_GKBA}
 &\langle\hat X_{12}(t)\hat X^\dagger_{34}(t')\rangle \approx  
 \nonumber \\ &
 i\sum_{M_1,M_2}\left[
 \tilde{\mathcal{G}}{}^r_{12,M_1M_2}(t-t')
 \langle\hat X_{M_1M_2}(t')\,\hat X^\dagger_{34}(t')\rangle 
 \right. \\ & \left. \qquad\quad
 - \langle \hat X_{12}(t)\,\hat X^\dagger_{M_1M_2}(t)\rangle
 \tilde{\mathcal{G}}{}^a_{M_1M_2,34}(t-t')
 \right]
 \nonumber
\end{align} 
and a similar expression for 
$\langle\hat X^\dagger_{34}(t')\, \hat X_{12}(t) \rangle$. 
Here we introduced the retarded and advanced Green functions on 
the Keldysh anti-contour\cite{Banyai}
\begin{align}
 \label{Gr_akb}
 \tilde{\mathcal{G}}{}^r_{12,34}(t) &\equiv -i\theta(-t) 
 \ll \hat X_{21}\hat I_B \rvert e^{-i\mathcal{L}t}\rvert\hat X_{43}
 \hat\rho_B^{eq}\gg 
 \nonumber \\
 &\equiv -i\theta(-t)\ll\hat X_{21}\rvert e^{-i\ {}^\theta\mathcal{L}_{eff}t}
  \rvert\hat X_{43}\gg
 \\
 \label{Ga_akb}
 \tilde{\mathcal{G}}{}^a_{12,34}(t) &\equiv i\theta(t) 
 \ll\hat X_{34}\hat I_B \rvert e^{i\mathcal{L}t}\rvert\hat X_{12}
 \hat\rho_B^{eq}\gg 
 \nonumber \\
 &\equiv i\theta(t) \ll\hat X_{34}\rvert e^{i\ {}^\theta\mathcal{L}_{eff}t}
  \lvert \hat X_{12}\gg
\end{align}
where ${}^\theta\mathcal{L}_{eff}$ is the effective Liouvillian generating 
the time-reversed evolution. Its connection to the effective Liouvillian 
$\mathcal{L}_{eff}$ generating forward-time evolution is\cite{SJ1,SJ2}
\begin{equation}
 \label{Lconnect}
 {}^\theta\mathcal{L}=\mathcal{L}^{*}
\end{equation}
Note difference in sign with Refs.~\onlinecite{SJ1,SJ2} which is due to our definition
of evolution operator as $e^{-i\ {}^\theta\mathcal{L}_{eff}t}$
instead of $e^{{}^\theta\mathcal{L}_{eff}t}$ in Refs.~\onlinecite{SJ1,SJ2}.

Applying (\ref{anti_GKBA}) to (\ref{EOM}) leads to a time-local
version of the generalized QME
\begin{equation}
 \label{sGQME}
 \frac{d}{dt}\sigma_{12}(t) = \sum_{3,4}\mathcal{L}_{12,34}\, \sigma_{34}(t)
\end{equation}
where in the molecular eigenbasis
\begin{align}
 \label{LGQME}
 &\mathcal{L}_{12,34} = -i(E_1-E_2)\,\delta_{1,3}\,\delta_{2,4}
 -\sum_\mathcal{M}\int_{-\infty}^{+\infty}dt
 \nonumber \\ & \left\{
 \  (-1)^{N_1-N_2}\,\delta_{N_1-1,N_3}\,\delta_{N_2-1,N_4}\times
 \right. \nonumber \\ &
 \left[\tilde{\mathcal{G}}{}^a_{(3,1)\mathcal{M}}(t)\,
 \Sigma^{<}_{\mathcal{M}(4,2)}(-t) - \Sigma^{<}_{(3,1)\mathcal{M}}(t)\,
 \tilde{\mathcal{G}}{}^r_{\mathcal{M}(4,2)}(-t)\right]
 \nonumber \\ &
 -(-1)^{N_1-N_2}\,\delta_{N_1+1,N_3}\,\delta_{N_2+1,N_4}\times
 \nonumber \\ &
 \left[\tilde{\mathcal{G}}{}^a_{(2,4)\mathcal{M}}(t)\,
 \Sigma^{>}_{\mathcal{M}(1,3)}(-t) - \Sigma^{>}_{(2,4)\mathcal{M}}(t)\,
 \tilde{\mathcal{G}}{}^r_{\mathcal{M}(1,3)}(-t)\right]
 \nonumber \\ &
 +\delta_{1,3}\,\delta_{N_2,N_4}\sum_s\left[
 \tilde{\mathcal{G}}{}^a_{(N_2-1\, s,4)\mathcal{M}}(t)\,
 \Sigma^{>}_{\mathcal{M}(N_2-1\, s,2)}(-t) 
 \right. \nonumber \\ &\left.\qquad\qquad + 
 \Sigma^{<}_{(2,N_2+1\, s)\mathcal{M}}(t)\,
 \tilde{\mathcal{G}}{}^r_{\mathcal{M}(4,N_2+1\, s)}(-t)\right]
 \nonumber \\ &
 -\delta_{2,4}\,\delta_{N_1,N_3}\sum_s \left[
 \tilde{\mathcal{G}}{}^a_{(3,N_1+1\, s)\mathcal{M}}(t)\,
 \Sigma^{<}_{\mathcal{M}(1,N_1+1\, s)}(-t) 
 \right. \nonumber \\ &\left.\left. \qquad\qquad +  
 \Sigma^{>}_{(N_1-1\, s,1)\mathcal{M}}(t)\,
 \tilde{\mathcal{G}}{}^r_{\mathcal{M}(N_1-1\, s,3)}(-t)
 \right] \right\}
\end{align}
Note that this expression for the effective Liouvillian in an arbitrary 
basis will differ from Eq.(\ref{LGQME}) only in the free evolution term.

If the QME, Eqs.~(\ref{sGQME}) and (\ref{LGQME}), is used with the free
molecular evolution instead of the effective dynamics in Eqs.~(\ref{Gr_akb})
and (\ref{Ga_akb}), one gets the standard Markovian Redfield QME.
The RWA is justified when $\Gamma_{mn}\ll|\varepsilon_a-\varepsilon_b|$
($m,n\in\{a,b\}$) in the molecular eigenbasis and consist in neglecting
the nondiagonal of $\Gamma_{mn}$ (here the standard effective single orbital
formulation is used)\cite{Esposito06}. As a result,
coherences become decoupled from population in the molecular
eigenbasis. The former die off at steady state and later obey a
Pauli rate equation. When the RWA is not justified, coherence effects
in the molecular eigenbasis cannot be neglected.

It should be clear from our derivation that time-local
form Eq.(\ref{sGQME}) of the generalized QME should have a similar degree of 
accuracy as the time-nonlocal version proposed earlier.\cite{EspoGalpPRB}
However, the present form is more suitable for realistic calculations, but 
most important it naturally suggests a self-consistent procedure to 
evaluate the effective Liouvillian. Indeed, the effective Liouvillian, 
Eq.(\ref{LGQME}), depends on the Liouville space Green functions, 
Eqs.~(\ref{Gr_akb}) and (\ref{Ga_akb}), which in turn depend on the 
Liouvillian through the connection (\ref{Lconnect}).


\subsection*{Expression for current}
The general expression for the time-dependent current at the contact-molecule 
interface $K$ can be calculated within the non-equilibrium Hubbard Green function 
approach as\cite{JauhoMeirWingreen,Hubbard}
\begin{align}
 I_K(t) &= 2\,\mbox{Im}\int_{-\infty}^t dt'\,\sum_{\mathcal{M},\mathcal{M}'}  
 \nonumber \\ &
 \left[ \ 
 \langle\hat X_\mathcal{M}(t)\,\hat X^\dagger_{\mathcal{M}'}(t')\rangle
 \Sigma^{<\, K}_{\mathcal{M}'\mathcal{M}}(t'-t) 
 \right. \\ & \left. +
 \langle\hat X^\dagger_{\mathcal{M}'}(t')\,\hat X_\mathcal{M}(t)\rangle
 \Sigma^{>\, K}_{\mathcal{M}'\mathcal{M}}(t'-t)
 \right]
 \nonumber
\end{align}
Using our ansatz, Eq.(\ref{anti_GKBA}), leads to the expression
\begin{align}
 \label{I_anti_GKBA}
 & I_K(t) = -2\,\mbox{Re}\int_{-\infty}^{+\infty}dt'
 \sum_{\mathcal{M},\mathcal{M}'}\sum_s
 \\ & \left[
 \sigma_{N\, s,N\, s_f}(t)\,
 \tilde{\mathcal{G}}{}^a_{(N\, s,N+1\, s_i)\mathcal{M}'}(t-t')\,
 \Sigma^{<\, K}_{\mathcal{M}'\mathcal{M}}(t'-t)
 \right. \nonumber \\ & \left.
 +
 \sigma_{N+1\, s_i,N+1\, s}(t)\,
 \tilde{\mathcal{G}}{}^a_{(N\, s_f,N+1\, s)\mathcal{M}'}(t-t')\,
 \Sigma^{>\, K}_{\mathcal{M}'\mathcal{M}}(t'-t)
 \right]
 \nonumber
\end{align}
where $\mathcal{M}\equiv (N\, s_f,N+1\, s_i)$.


\subsection*{Full counting statistics}
The theory of full counting statistics (FCS) was initially proposed by 
Levitov and Lesovik,\cite{LevitovLesovik1,LevitovLesovik2} and became popular 
in the molecular electronics community when shot noise in molecular junctions 
became experimentally measurable.\cite{Ruitenbeek} The theoretical formalism 
for FCS within the QME was developed in Ref.~\onlinecite{EspositoReview}. 
Within the FCS the evolution operator is dressed by counting field(s) 
$\lambda$ which track the exchange of electrons between the molecule and 
the contact(s)
\begin{equation}
 \hat V^{\lambda}_{KM} = \sum_{k\in K}\sum_\mathcal{M}\left(
 V_{\mathcal{M}k}e^{i\lambda/2}\hat X_\mathcal{M}^\dagger\hat c_k + 
 V_{k\mathcal{M}}e^{-i\lambda/2}\hat c_k^\dagger\hat X_\mathcal{M} \right)
\end{equation}
Below we assume that counting starts at time $t_0$.

The dressed Liouville equation for the total density matrix takes the form
\begin{equation}
 |\hat\rho_\lambda(t)\gg=\exp[-i\mathcal{L}_\lambda(t-t_0)]\,|\hat\rho(t_0)\gg
\end{equation}
As usual initial condition is assumed to be a direct product of the molecular and bath density matrices
\begin{equation}
 |\hat\rho(t_0)\gg = |\hat\sigma(t_0)\,\hat\rho_B^{eq}\gg
\end{equation}
so that dressed evolution of the system DM becomes
\begin{equation}
 |\hat\sigma_\lambda(t)\gg = \exp[-i\mathcal{L}_{eff,\lambda}(t-t_0)]\,
 |\hat\sigma(t_0)\gg
\end{equation}

The FCS is given by the generating function
\begin{equation}
 G(t,\lambda) \equiv \ll\hat I|\hat\rho_\lambda(t)\gg
\end{equation}
The long time limit of the logarithm of the generating function,
\begin{equation}
 S(\lambda) \equiv \lim_{t\to\infty} \frac{1}{t}\ln G(t,\lambda),
\end{equation}
provides information on the steady-state cumulants of the FCS
\begin{equation}
 C_n\equiv\frac{d^n}{d(i\lambda)^n}S(\lambda)
\end{equation}
In particular, the steady-state current is given by the first cumulant and the second cumulant yields to the zero-frequency shot noise. Using the spectral decomposition of the effective Liouvillian
\begin{equation}
 \mathcal{L}_{eff,\lambda} = \sum_\gamma |R_\gamma(\lambda)\gg\,
 \nu_\gamma(\lambda)\,\ll L_\gamma(\lambda)|
\end{equation}
and since in the long time limit only one eigenmode $\nu_0(\lambda)$ survives, 
we get
\begin{align}
 \label{I_FCS}
 I &= -\frac{d}{d\lambda}\nu_0(\lambda)
 \\
 \label{S_FCS}
 S &= i\frac{d^2}{d\lambda^2}\nu_0(\lambda)
\end{align}
We note that both (\ref{I_anti_GKBA}) and (\ref{I_FCS}) provide the same 
steady-state current.


\section{\label{num}Numerical examples}
We now compare the results predicted by our new self-consistent approach with the standard Markovian Redfield equation and the nonequilibrium Green function results. The simplest model which can be treated by these three methods, while providing information on both populations and coherences, is a non-interacting two-level bridge (TLB) between metallic contacts. 

The bridge part of the model has 4 many-body states: 
$\lvert 0\rangle\equiv\lvert 0,0\rangle$, 
$\lvert a\rangle\equiv\lvert 1,0\rangle$,
$\lvert b\rangle\equiv\lvert 0,1\rangle$, and
$\lvert 2\rangle\equiv\lvert 1,1\rangle$, 
where $\lvert n_a,n_b\rangle$ indicates number of electrons 
$n_{a,b}=\{0,1\}$ on the level $a$ and $b$, respectively. 
The relevant single electron transitions $\mathcal{M}$ are: 
$(0,a)$, $(0,b)$, $(b,2)$, and $(a,2)$. 
They are connected to the second quantized (single-particle) 
excitation operators used in standard GF approaches by
\begin{align}
 \label{da}
 \hat d_a &= \hat X_{0a} + \hat X_{b2} \\
 \label{db}
 \hat d_b &= \hat X_{0b} - \hat X_{a2}
\end{align}
where $\hat d^\dagger_{a,b}$ ($\hat d_{a,b}$) are the creation (annihilation) operators for an electron on level $a$ and $b$, respectively.

The molecular Hamiltonian, Eq.(\ref{HM}), in this case takes the form
\begin{equation}
 \label{HM_TLB}
 \hat H_M = \varepsilon_a\hat X_{aa} + \varepsilon_b\hat X_{bb}
          + t\left(\hat X_{ab}+\hat X_{ba}\right)
\end{equation}
The model was treated extensively within the QME approach.\cite{Novotny02,Esposito06,Esposito07,Esposito08}

The TLB model is easily exactly solved using NEGF. The connection, Eqs.~(\ref{da}) and (\ref{db}), provides partial information about the many-body state populations but full information about coherences, which makes the comparison between the different approaches meaningful. In particular,
\begin{align}
 \label{Gaa}
 -iG^{<}_{aa}(t,t) &= \sigma_{00}(t) + \sigma_{bb}(t) \\
 \label{Gbb}
 -iG^{<}_{bb}(t,t) &= \sigma_{00}(t) + \sigma_{aa}(t) \\
 \label{Gab}
 -iG^{<}_{ab}(t,t) &= \sigma_{ab}(t)
\end{align}

Inter-level coherence in the TLB is either due to hoping matrix element $t$ or due to the coupling of the two levels via a common bath. The latter interference enters through the non-diagonal elements of the bridge-contact coupling matrix
\begin{align}
 \Gamma^K_{ab} &\equiv \sum_{k\in K} V_{ak}V_{kb}\delta(E-\varepsilon_k) 
 \nonumber \\
 &\equiv \ \sum_{k\in K} V_{(oa)k}V_{k(0b)}\delta(E-\varepsilon_k) 
 \\
 &\equiv - \sum_{k\in K} V_{(b2)k}V_{k(a2)}\delta(E-\varepsilon_k) 
\end{align}
where $K=L,R$. The matrix is energy independent in the wide-band limit assumed below. We intentionally perform simulations at low temperature $T=10$~K to avoid broadening of the bridge levels due to artificially high values of the temperature. We focus on the relevant temperature regime for molecular junctions: $\Gamma \gg k_BT$.

\begin{figure}[htbp]
\centering\includegraphics[width=\linewidth]{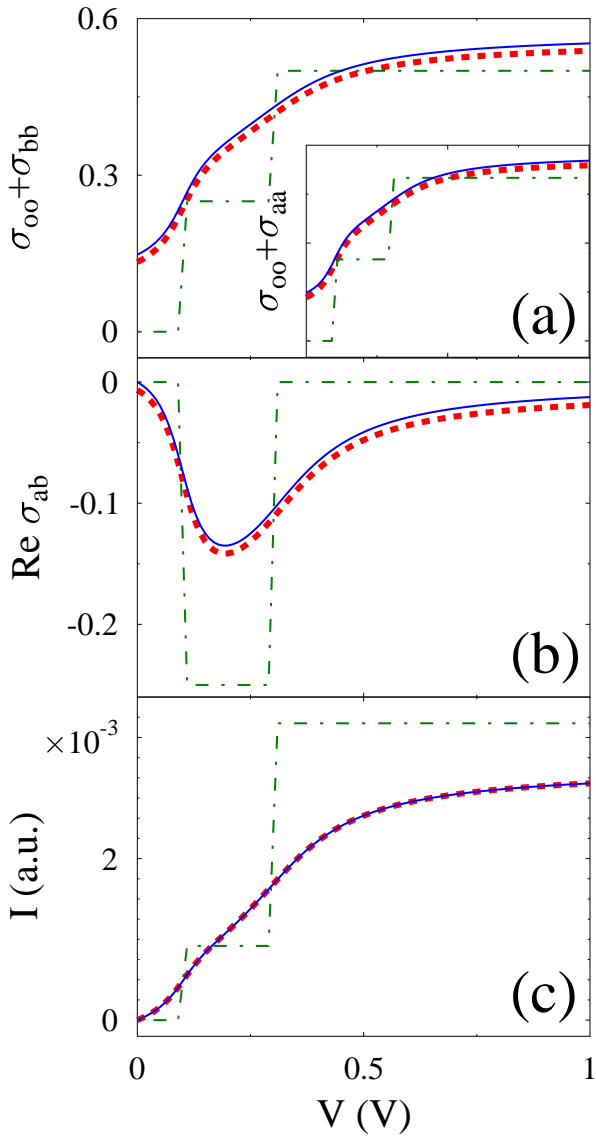}
\caption{\label{f1}
(Color online)
Two-level bridge by Redfield QME (dash-dotted line, green), NEGF (dashed line, red), and self-consistent GQME (solid line, red) approaches. Shown are 
(a) probabilities, (b) coherence, and (c) current vs. bias. Far off-resonant local basis treatment. See text for parameters.
}
\end{figure}

We start our consideration from a simple far off-resonant case, 
where $\Gamma_{mn}\ll|\varepsilon_a-\varepsilon_b|$ ($m,n\in\{a,b\}$). 
Here our consideration is done in a local molecular basis ($t\neq 0$). 
The parameters of the calculation are the following: 
level positions $\varepsilon_a=0.2$~eV and $\varepsilon_b=0.5$~eV, 
inter-level hopping $t=0.1$, escape rates for both levels into both contacts 
$\Gamma^K_{aa}=\Gamma^K_{bb}=0.1$~eV ($K=L,R$), 
and level mixing due to coupling to the contacts 
$\Gamma^K_{ab}=\Gamma^K_{ba}=0.05$~eV. 
The Fermi energy in the absence of bias is taken as zero, 
$E_F=0$, and the voltage division factor is $1$, 
i.e. $\mu_L=E_F+|e|V$ and $\mu_R=E_F$. 
The NEGF calculation is performed on an energy grid spanning the range from 
$-10$ to $10$~eV with step $10^{-4}$~eV. Both the Redfield QME and our scheme 
are expected to work properly in this region of parameters. 
Figure~\ref{f1} compares the results of the Redfield QME and our new scheme 
to the exact NEGF results for the model. The main graph and the inset in 
Fig.\ref{f1}a display populations, Eqs.~(\ref{Gaa}) and (\ref{Gbb}). 
The real part of the coherence, Eq.(\ref{Gab}), is shown in Fig.\ref{f1}b. 
The imaginary part of the coherence is zero in this case (not shown). 
As discussed earlier,\cite{EspoGalpPRB} the Redfield QME approach misses 
the information about the broadening of the molecular levels. 
Our approach accurately recovers this information. Both populations and 
coherence are well reproduced by our GQME as well as the current-voltage 
characteristics (see Fig.\ref{f1}c). Qualitatively, the predictions of 
the Redfield QME approach are also correct here.

\begin{figure}[htbp]
\centering\includegraphics[width=\linewidth]{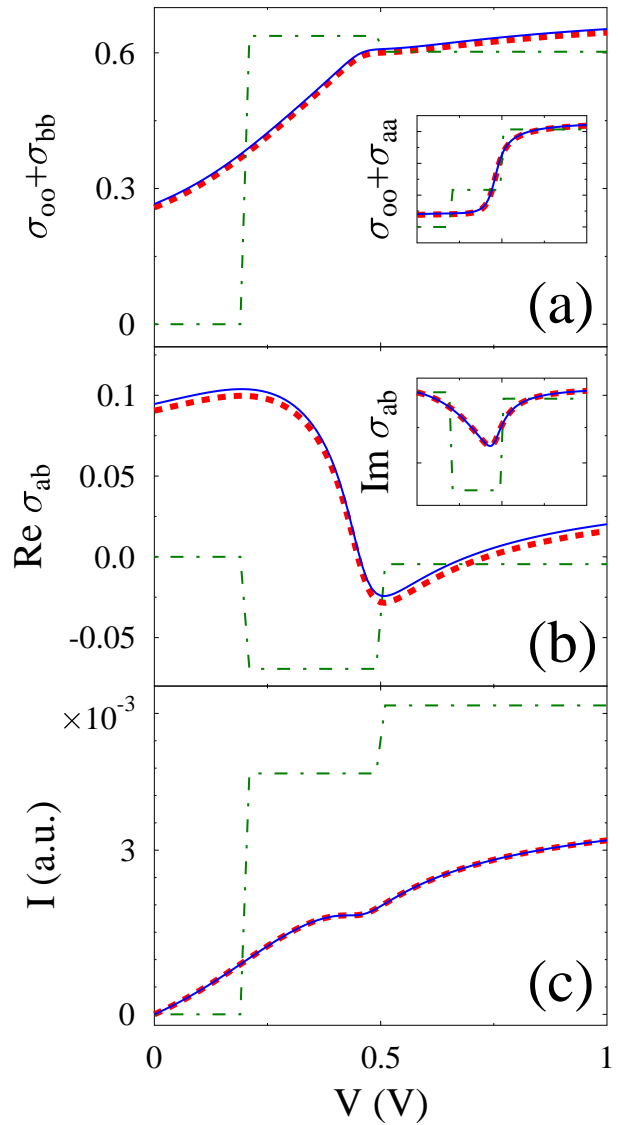}
\caption{\label{f2}
(Color online)
Two-level bridge by Redfield QME (dash-dotted line, green), 
NEGF (dashed line, red), and self-consistent GQME (solid line, red) approaches. 
Shown are (a) probabilities, (b) coherence, and (c) current vs. bias. 
Case of strong mixing due to coupling to the contact(s).
See text for parameters.
}
\end{figure}

We next consider the case of strong mixing due to coupling to the contact(s).
Parameters of the calculation are  $t=0$, 
$\Gamma^L_{aa}=0.3$~eV, $\Gamma^R_{aa}=0.048$~eV,
$\Gamma^L_{bb}=0.2$~eV, $\Gamma^R_{bb}=0.1125$~eV,
$\Gamma^L_{ab}=\Gamma^L_{ba}=0.12$~eV, and
$\Gamma^R_{ab}=\Gamma^R_{ba}=0.15$~eV. 
The other parameters are the same as in Fig.\ref{f1}. 
Figure~\ref{f2} compares the Redfield QME and our self-consistent GQME scheme 
with the exact NEGF results. Our scheme still reproduces populations 
(Fig.\ref{f2}a), coherences (Fig.\ref{f2}b), 
and current (Fig.\ref{f2}c) accurately. 
The Redfield QME however fails to reproduce, even qualitatively, 
the populations (see Fig.\ref{f2}a) and the real part of the coherence 
(see Fig.\ref{f2}b). This is due to coherence effects through the bath
which are not properly captured by the Redfield QME.

\begin{figure}[htbp]
\centering\includegraphics[width=\linewidth]{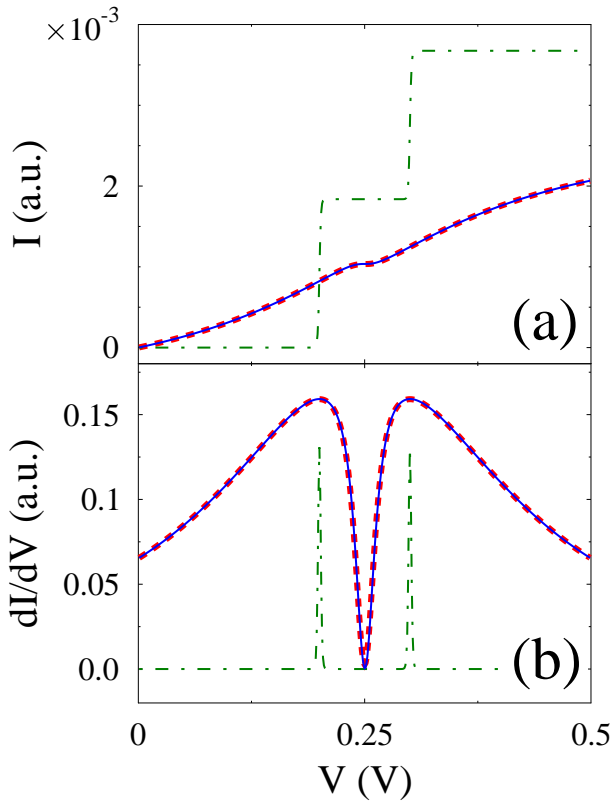}
\caption{\label{f3}
(Color online)
Two-level bridge by Redfield QME (dash-dotted line, green), NEGF (dashed line, red), and self-consistent GQME (solid line, red) approaches. Shown are (a) current and (b) conductance vs. bias. Resonant, $\varepsilon_a=\varepsilon_b$, consideration in a local basis with only level $a$ coupled to contacts, $\Gamma^K_{aa}\neq 0$ and $\Gamma^{K}_{bb}=\Gamma^K_{ab}=\Gamma^K_{ba}=0$ ($K=L,R)$. See text for parameters.
}
\end{figure}

Interference effects in molecular systems were observed experimentally
for electron transfer\cite{Joachim} and molecular junction 
currents\cite{MayorWeber} involving derivatives of benzene connected 
in meta or para position. 
They were also extensively discussed in theoretical literature.\cite{Mazumdar07,RatnerJACS07,RatnerJACS08com,RatnerJACS08,RatnerJCP08,RatnerCPC09} 
Figure~\ref{f3} shows current (a) and conductance (b) vs. bias, 
for a model system in which destructive interferences can be experimentally 
observed and which has been discussed earlier in the literature.\cite{RatnerCPC09} 
This model is a two-level system with only one of the levels, $a$, 
attached to both $L$ and $R$ contacts. The other level, $b$, 
is coupled to the level $a$ through a hopping element $t$. 
Level $b$ is not directly attached to any contacts. 
As a result, the tunneling electron has two possible pathways
to be transferred from contact $L$ to contact $R$ via the system: 
one directly through level $a$ and the other by exploring level $b$ on its way. 
Interference between the two paths leads to destructive interference in 
the transport characteristics, which reveals itself as a dip in the conductance.
Parameters of the calculation are $\varepsilon_a=\varepsilon_b=0.25$~eV, 
$t=0.05$~eV, $\Gamma^L_{aa}=\Gamma^R_{aa}=0.2$~eV 
and $\Gamma^K_{ab}=\Gamma^K_{ba}=\Gamma^K_{bb}=0$. 
Fig.\ref{f3}b shows that destructive interference in conductance is accurately 
captured by our self-consistent GQME approach but not by the Redfield QME. 

\begin{figure}[htbp]
\centering\includegraphics[width=\linewidth]{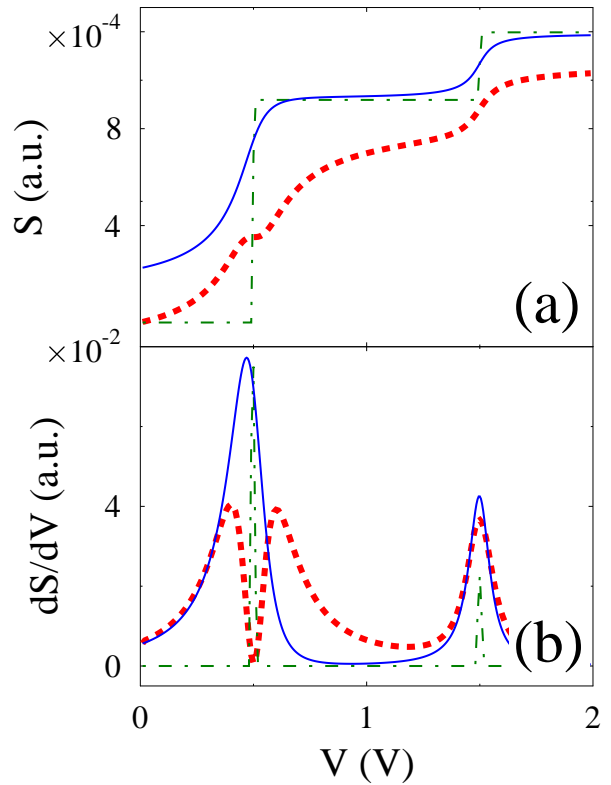}
\caption{\label{f4}
(Color online)
Two-level bridge by Redfield QME (dash-dotted line, green), NEGF (dashed line, red), and self-consistent GQME (solid line, red) approaches. Shown are (a) shot noise and (b) differential shot noise vs. bias. See text for parameters.
}
\end{figure}

Finally, we present the results of calculation for the zero-frequency shot noise. The shot noise was calculated using the FCS approach, which for the Redfield QME and our self-consistent scheme yields to expression (\ref{S_FCS}). The FCS within NEGF was discussed in our previous publication.\cite{spin_pump} 
Note that the FCS for a two-level bridge within QME was also discussed in Ref.~\onlinecite{Esposito07} 
We intentionally consider two independent levels 
(no coupling either within the system or through the bath)
to demonstrate the general problem that usual QME schemes 
have in representing higher moments. 
We assume that the level $\varepsilon_a=0.5$~eV is coupled 
symmetrically to both contacts $\Gamma^L_{aa}=\Gamma^R_{aa}=0.1$~eV. 
The other level $\varepsilon_b=1.5$~eV is coupled asymmetrically
$\Gamma^L_{bb}=10\Gamma^R_{bb}=0.1$~eV. 
Other parameters are $t=0$, $\Gamma^K_{ab}=\Gamma^K_{ba}=0$, and $T=10$~K.

Figure~\ref{f4} compares the results of the Redfield QME and the self-consistent GQME
with the exact results provided by NEGF. It is known\cite{our_noise} that 
the differential shot noise yields a two-peak structure for a symmetrically 
coupled molecule, with only a single peak observed for highly asymmetric 
coupling. One sees that the latter peak (centered around $1.5$~eV in Fig.\ref{f4}b)
is reproduced quite well by our self-consistent approach. 
The double well structure is however missed by both the Redfield QME 
and our approach (see peak centered aroundi $0.5$~eV). 
Cumulants beyond first order can be very sensible to approximations and our 
level of theory, which neglects molecule-contacts correlations beyond second 
order, has to be improved to properly reproduce them.  

Summarizing, within our simple two-level model of molecular junction, 
the comparison of the proposed self-consistent GQME and 
the Markovian Redfield QME to the exact results provided by the NEGF 
shows that the self-consistent GQME is highly accurate even when 
the Redfield QME fails qualitatively. We find that the self-consistent GQME 
will only fail in the region of parameters where the inter-level coherence 
due to the coupling to contacts in the molecular eigenbasis is bigger than 
the distance between the molecular eigenstates and is of the order of 
the molecule-contact coupling strength (i.e. exactly at resonance). 
As a result, in most relevant practical calculations of transport in 
molecular junctions, we expect our approach to be accurate. 
The ability of the proposed scheme to treat molecular transport in 
the language of many-body states of the isolated molecule and its ability 
to properly account for interference effects makes it a valuable practical 
tool for ab initio calculations of transport in molecular junctions. 


\section{\label{conclude}Conclusion}
We present a practical scheme for molecular transport calculation.
The scheme is based on a time-local generalized quantum master equation
obtained by closing the exact EOM for Hubbard operators by employing
a time-reversed evolution ansatz on the Keldysh anti-contour,
similar to generalized Kadanoff-Baym ansatz introduced in our previous
publication. We note that the approximations involved in derivation of
the time-local equation are essentially the same as for the earlier
(more traditional) time-nonlocal GQME. The time-locality of the GQME
allows us to formulate a feasible self-consistent scheme to calculate
the time dependent molecular density matrix and the current in term of
molecular many-body states. We find that the convergence of
the self consistent method for a simple two-level bridge model
is achieved within two iterative steps. The results of the calculation
for TLB within self-consistent GQME are compared to Redfield QME approach,
and to exact NEGF results. We demonstrate that our scheme
(contrary to the usual QME result) properly captures populations and coherences.
In particular, destructive interference effects in the molecular devices
previously discussed in the literature can now be properly described in
the many-body states language, which makes the scheme a valuable tool
for practical ab initio calculations. Current-voltage characteristics
are reproduced with high accuracy. We are able to calculate the molecular
device characteristics in the experimentally relevant regime of
low temperatures. We find that the scheme breaks down in the region of
the parameters where coherences in the system eigenbasis
(i.e. coherences introduced through non-diagonal elements of molecule-contact
coupling matrix $\Gamma$) are both bigger than the inter-level separation
and are of the order of the escape rates (diagonal elements of
the molecule-contact coupling matrix $\Gamma$).
Thus, we expect the scheme to be a valuable tool for most practical transport calculations in molecular junctions.

The formulation of a scheme capable of reproducing higher moments of the full counting statistics in the many-body state language that goes beyond the 
Redfield QME approach\cite{EspositoReview} is the goal of future research.


\begin{acknowledgments}
M.E. is supported by the Belgian Federal Government (IAP project ``NOSY").
M.G. gratefully acknowledges support of the UCSD (Startup Fund) and US-Israel Binational Science Foundation.
\end{acknowledgments}

\bibliography{akb}


\end{document}